\documentclass[aps,prl,reprint, amsmath, amssymb,superscriptaddress]{revtex4-1}
 
\usepackage{bm}
\usepackage[retainorgcmds]{IEEEtrantools}
\usepackage{graphicx}
\usepackage{mathrsfs}
\usepackage[mathscr]{eucal}
\usepackage{amsmath}
\usepackage{amssymb}
\usepackage{color}
\usepackage{amsfonts}
\usepackage{amsthm}
\usepackage{times,txfonts}
\usepackage{enumerate}
\usepackage{dsfont}
\usepackage{nicefrac}
\usepackage[colorlinks=true,linkcolor=blue,urlcolor=blue,citecolor=blue,pdfusetitle]{hyperref}
\hypersetup{breaklinks=true}
\definecolor{darkgreen}{rgb}{0.2, 0.3, 0.1}

\DeclareMathOperator{\tr}{tr}

\newlength{\eqboxstorage}

\begin{document}

\title{Universal many-body diffusion from momentum dephasing}
\date{\today}
\author{Maur\'icio Hippert}
\email{hippert@illinois.edu}
\affiliation{Illinois Center for Advanced Studies of the Universe\\ Department of Physics, 
University of Illinois at Urbana-Champaign, Urbana, IL 61801, USA}
\author{Gabriel T. Landi}
\email{gtlandi@gmail.com}
\affiliation{Instituto de F\'isica da Universidade de S\~ao Paulo, 05314-970 S\~ao Paulo, Brazil}
\author{Jorge Noronha}
\email{jn0508@illinois.edu}
\affiliation{Illinois Center for Advanced Studies of the Universe\\ Department of Physics, 
University of Illinois at Urbana-Champaign, Urbana, IL 61801, USA}

\begin{abstract}
The open dynamics of quantum many-body systems involve  not only the exchange of energy, but also of other conserved quantities, such as momentum. 
This leads to additional decoherence, which may have a profound impact in the dynamics.
Motivated by this, we consider a many-body system subject to total momentum dephasing and show that under very general conditions this leads to a diffusive component in the dynamics of any local density, even far from equilibrium. 
Such component will usually have an intricate interplay with the unitary dynamics. 
To illustrate this, we consider the case of a superfluid and show that momentum dephasing introduces a damping in the sound-wave dispersion relation, similar to that predicted by the Navier-Stokes equation for ordinary fluids.
Finally, we also study the effects of dephasing in linear response, and show that it leads to a universal additive contribution to the diffusion constant, which can be obtained from a Kubo formula.

\end{abstract}

\maketitle

\emph{\textbf{Introduction --}} Deriving emergent collective behavior from the underlying microscopic motion is a century-old problem in physics that continues to be actively pursued. A scenario of particular importance is the emergence of diffusive behavior in quantum many-body systems. 
Examples include  superfluids or ultracold gases in optical lattices~\cite{Greiner2002,Bloch2008,Gring2012,Brantut2012,Langen2013,Langen2015,Kaufman2016}, where
non-integrability is often credited with inducing diffusive transport~\cite{Bertini2020b}. Another example stems from the quark-gluon plasma formed in relativistic heavy-ion collisions \cite{Heinz:2013th}, where the shear viscosity to entropy density ratio is found \cite{Bernhard:2019bmu} to nearly saturate estimates based on the uncertainty principle \cite{Danielewicz:1984ww,Kovtun:2004de}.  

Diffusive behavior describes how gradients drive the universal late time evolution of perturbations in a conserved density at long wavelengths \cite{Kadanoff1963}. In fact, the evolution of conserved quantities becomes arbitrarily slow at sufficient long wavelengths and, as a consequence, the corresponding hydrodynamic modes control the late-time relaxation of perturbations. Explicit expressions to describe the approach to equilibrium follow from a gradient expansion of the non-equilibrium currents which, for diffusion, leads to Fick's law and higher-order gradient corrections \cite{forster1975hydrodynamic}. In the case of unitary microscopic dynamics,  quantities such as the diffusion coefficient that defines Fick's law can be computed from suitably defined Kubo formulas \cite{doi:10.1143/JPSJ.12.570} involving the long time, long wavelength limit of two-point functions. 
In certain models, such coefficients can also be analytically derived from nonlinear interactions, e.g., in some classical anharmonic lattices \cite{Aoki2000a,Aoki2006}.  

Recent foundational results~\cite{Deutsch1991,Srednicki1994,Popescu2006,Goldstein2006a} and experiments~\cite{Gring2012,Brantut2012,Langen2013,Langen2015,Kaufman2016} show that many-body systems in non-thermal quantum states can still exhibit thermal properties. This raises the question of whether hydrodynamic or diffusive behavior can also emerge in such states. 
An alternative route  to  quantum diffusion is through decoherence, for instance, via bulk noises such as dephasing~\cite{Palma1996,breuer2002theory}.
Generally speaking, these represent reservoirs which do not exchange excitations with the system, but notwithstanding still inject noise and hence interfere with the transport properties. 
In fact, in integrable systems arbitrarily small dephasing strengths are known to turn the transport diffusive~\cite{Asadian2013,Znidaric2010a,landi2021nonequilibrium,Malouf2018}.
Remarkably, this is also the case in the presence of disorder~\cite{Znidaric2013a,Znidaric2017}. Furthermore, the analog of dephasing can also be implemented in classical systems, as a form of energy-conserving noise~\cite{Bolsterli1970,Bernardin2011a,Dhar2011,Bonetto2008,Guimaraes2015}, which leads to diffusion in both integrable and non-integrable models. 

A crucial feature concerning open many-body quantum systems is the decoherence process induced by charge conservation. 
The interaction between the system and its surroundings can cause not only the exchange of energy, but also of other conserved quantities~\cite{Vaccaro2011,Langen2015,Guryanova2016,YungerHalpern2018}.
If a system exchanges a conserved quantity $Q$ with the environment, this leads to quantum correlations (entanglement or discord) between them, as the system gains exactly the same amount of $Q$ that the environment has lost~\cite{Brandao2013,Horodecki2013,Guryanova2016,Manzano2020}. 
As a consequence, coherences on the basis of conserved quantities are attenuated~\cite{Zurek1981,Walls1985a}. 
Since decoherence is already known to play an important role in diffusion, decoherence driven by charge conservation should thus affect transport phenomena in a very large class of quantum many-body systems. In particular, it should also alter transport coefficients.

In this Letter we consider generic open quantum many-body systems subject to a quantum master equation describing dephasing of the total momentum, which is one of the conserved charges a system may naturally exchange with the environment.
Under very general conditions, we show that this type of dephasing leads to diffusive behavior, evidenced by the emergence of a diffusion equation for the local density, which can occur even in states that are arbitrarily far from thermodynamic equilibrium. 
We also analyze the case of a superfluid and show that momentum dephasing leads to dissipative effects that damp sound waves in a way that is similar to what happens in ordinary fluids described by the Navier-Stokes equations.
Finally, we  address the effect of decoherence in the context of linear response, close to equilibrium. 
We show that the contribution of the momentum dephasing to the diffusion coefficient is both additive and universal. We argue that similar contributions should also arise in other transport coefficients.

\emph{\textbf{Momentum dephasing dynamics --}}
Consider a quantum many-body system which evolves in contact with an environment. 
We assume the effects of the environment are Markovian, so that the reduced density matrix $\rho$ of the system may be described by a
Lindblad
master equation~\cite{Lindblad1976,Gorini1976}:
\begin{equation}
 \frac{d{\rho}(t) }{dt} = \mathcal{L}(\rho) =-\frac{i}{\hbar}\, [\mathcal H, {\rho}(t)] +\mathcal{D}({\rho}(t))\,,
 \label{eq:unperturbedLind}
\end{equation}
where  $\mathcal{H}$ is the system Hamiltonian and  $\mathcal{D}$ is the dissipator, which encodes the non-unitary, but probability-conserving, part of the evolution. 
The dissipator can be written as
\begin{equation}
 \mathcal{D}\left({\rho}\right)=\sum_k \gamma_k \left(L_k {\rho} L_k^\dagger - \frac{1}{2}\{L_k^\dagger L_k, {\rho}\}\right)\,,
 \label{eq:dissipatorjump}
\end{equation}
with coefficients $\gamma_k\geq 0$ and an arbitrary set of operators $L_k$.

For concreteness, one may consider a system of $N$ interacting particles with positions $\bm x_a$ and momenta $\bm p_a$ \cite{Kadanoff1963}, though our considerations also apply to second quantized systems (as will be discussed below). The jump operators $L_k$ depend on the coupling to the environment, and will generally be given by combinations of $\bm x_a$ and $\bm p_a$ of all the particles.
Specifics of this coupling lead to very rich physics, and even novel phase transitions~\cite{Diehl2008}, but here we take a different route.
Universal hydrodynamic behavior stems from the long wavelength dynamics of conserved quantities, which is expected to be largely insensitive to details of the microscopic physics.  
Hence, we concentrate on the case of \emph{total momentum dephasing}, as this is a global conserved quantity ($[\mathcal{H},\bm P]=0$), for any interaction potential that depends only on interparticle distances, $V(\bm x_a - \bm x_b)$.
We thus assume that $L_k=L_k^\dagger=P_k$, $k =1,2,3$, where $\bm P=\sum_a \bm p_a$ is the total momentum of the system (or, equivalently, the momentum of the center of mass). 
This corresponds to a dephasing-type of dynamics.
For simplicity, we also assume that the coupling to the environment is isotropic,  $\gamma_{k} = \sigma^2$. 
Thus, we analyze Eq.~\eqref{eq:unperturbedLind} with 
\begin{equation}\label{eq:modelP}
\mathcal{D}({\rho}) = - \frac{\sigma^2}{2} \sum\limits_k [P_k, [P_k, {\rho}]].
\end{equation}
This simple choice is a minimal ingredient for the emergence of diffusive behavior, as shown below. 
While other jump operators, such as $\bm P^2$ could be considered, they correspond to higher orders in derivatives and should be subleading in the long-wavelength limit. 
This is also the case for jump operators constructed from the total momenta of smaller subsystems. 

Such a dissipator leads to the exponential damping of non-diagonal elements, or coherences, of the density matrix in the momentum representation.
For instance, if the unitary contribution in~\eqref{eq:unperturbedLind} can be neglected (e.g. $\mathcal{H}\simeq0$), one obtains $\langle \bm p|\rho|\bm p'\rangle \propto \exp(- \sigma^2(\bm p - \bm p')^2 t/(2\hbar))$, where $|\bm{p}\rangle$ is the eigenbasis of $\bm{P}$.  
Because $\bm P$ is the generator of spatial translations, Eq.~\eqref{eq:modelP} can be interpreted as the action of random, uncorrelated jumps in position of variance $\hbar^2\sigma^2 d t$ \cite{Wiseman2009}. 
The same dephasing also appears when the total momentum is continuously measured,  as will be discussed below.

\emph{\textbf{Emergence of diffusive behavior --}} 
From Eqs.~\eqref{eq:unperturbedLind} and \eqref{eq:modelP}, it follows that the average of any operator $\mathcal{A}$ evolves according to 
\begin{equation}\label{eq:dtAofx}
 \partial_t \langle \mathcal A \rangle (t) =   \tr\left\{ \mathcal A\, \partial_t {\rho}(t)\right\} 
= \frac{i}{\hbar} \langle  [\mathcal H, \mathcal A]\rangle (t)+ \langle {\mathcal{D}}(\mathcal A)\rangle(t)\,,
\end{equation}
where $\langle \mathcal A \rangle(t) := \tr\{\rho(t)\,\mathcal A\}$. 
In particular, we consider any local density operator,
which depends on position $\bm{x}$, and position operators $\bm x_a$, only through the difference $\bm{x} - \bm{x}_a$; this could be, e.g. the particle density 
$\mathcal N(\bm x) = \sum_a \delta(\bm x - \bm x_a)$~\cite{Kadanoff1963} 
(or the field operator $\Psi(\bm{x})$ in second quantization).
Since $\bm P$ is the generator of spatial translations, 
it follows that 
$\mathcal{D}(\mathcal N(\bm x)) =\hbar^2 \sigma^2 \nabla^2 \mathcal N(\bm x)/2$. 
Moreover, if the density is  a conserved quantity, it follows that $(i/\hbar)[\mathcal H, \mathcal{N}(\bm x)]=-\nabla \cdot \bm{\mathcal J}(\bm x)$, where $\bm {\mathcal J}$ is the associated current operator. 
Thus, Eq.~\eqref{eq:dtAofx} reduces to
\begin{equation}\label{eq:diffusion}
\partial_t n(t,\bm x) =  -\nabla \cdot \bm{j}(t,\bm{x}) + \frac{\hbar^2\sigma^2}{2} \nabla^2 n(t,\bm{x})\,,
\end{equation}
where $n(t,\bm x) :=\langle \mathcal N(\bm{x})\rangle(t)$, $\bm{j}(t,\bm x) :=\langle \bm{\mathcal J}(\bm{x})\rangle(t)$. 
The momentum dephasing thus leads to an exact diffusion-type term, in addition to the unitary contribution to the rhs of Eq.\ \eqref{eq:diffusion}. In general, both terms act together, mixing diffusive and coherent dynamics. 
In the particular case of very strong dephasing, the second term dominates and one finds a standard diffusion equation $\partial_t  n(t,\bm x) = D_{\textrm{deph}} \,\nabla^2 n(t,\bm x)$, where
\begin{equation}
D_{\textrm{deph}}= \frac{\hbar^2\sigma^2}{2}\,
\label{define_dephasing_D}
\end{equation}
is the effective diffusion constant created by momentum dephasing. Therefore, one can see that diffusive behavior can generally emerge in this scenario even in states arbitrarily far from equilibrium, given that no assumptions of near equilibrium behavior were made. This should be contrasted with textbook derivations of the diffusion equation in unitary many-body systems, which assume a Fick's law behavior $\bm j \sim - \nabla n$ in the long wavelength regime \cite{forster1975hydrodynamic}.

\emph{\textbf{Interpretation --}} Since the total momentum is conserved, quantum superpositions of momentum eigenstates are bound to loose coherence when the system exchanges momentum with the environment. Thus, the phenomena of diffusion from decoherence should be a ubiquitous feature of open quantum many-body systems.

A possible interpretation of Eq.~\eqref{eq:diffusion} is in terms of Heisenberg's uncertainty principle. Consider a scenario in which the total momentum of a many-body system is continuously measured \cite{Wiseman2009,Jacobs2014}. 
This can be achieved, for instance, through the sequential interaction of the system with $N_\theta$ probe particles, during successive time intervals of duration $\theta$.
Each probe particle is assumed to be described by a pair of canonical quadratures $q_i, p_i$ ($i =1,2,3$) satisfying $[q_i,p_j]=i\hbar\,\delta_{ij}$. 
The system-probe interaction is taken to be~\cite{Jacobs:IntroContMeas}  $\mathcal{H}_I =g\sum_i P_i\otimes q_i$, where $g$ denotes the coupling strength.
After each interaction, the probes are projectively measured in the eigenbasis of $p_i$, from which the system total momentum $P_i$ can be read as  $\langle P_i \rangle_\theta = \langle \Delta_\theta p_i\rangle/(g\theta)$, where $\Delta_\theta p_i$ is the shift in $p_i$ due to the interaction. %

Continuous measurements correspond to the limit where $g \theta$ is made arbitrarily small. The perturbation of the system due to each measurement is encoded in the evolution operator $U_\theta = e^{-i\,\mathcal H_I\,\theta/\hbar}$ \cite{breuer2002theory}. 
Assuming each probe is prepared in a state $\rho_\theta$, such that $\langle q_i \rangle=0$, expanding $U_\theta(\rho\otimes \rho_\theta) U_\theta^\dagger$ to second order in $\theta$ and averaging over the measurement outcomes yields Eqs.~\eqref{eq:unperturbedLind} and \eqref{eq:modelP}, with $\sigma^2 = g^2\theta (\delta q_i/\hbar)^2$, where $\delta q_i$ is the uncertainty in $q_i$. 
Although the uncertainty of each individual measurement becomes very large, the combined uncertainty decreases as $\delta P_i(t)= \delta p_i/\sqrt{g^2\theta\,t}$, where $\delta p_i$ is the uncertainty in the measurement of the probe. Moreover, applying the uncertainty principle to the probe, $\delta p_i \,\delta q_i\geq \hbar/2$, we find
\begin{equation}
  \left(\sqrt{2D_{\textrm{deph}}t} \,\right)^2\,(\delta P_i)^2 \geq \frac{\hbar^2}{4}\,.
\end{equation}
Thus, not only does the continuous measurement of the momentum give rise to a diffusion constant, but the diffusion is such that the uncertainty principle is always satisfied. 
This implies that if one tries to continuously track the direction of motion of the center of mass of a quantum many-body system, diffusion necessarily sets in. 
The universal diffusion constant in Eq.~\eqref{define_dephasing_D} follows from these simple considerations.

\emph{\textbf{Dephasing viscosity in superfluids --}}  
As a more complete example of a quantum many-body system, let us examine how the effects of momentum dephasing manifest in the case of a superfluid, described in second quantization by the Hamiltonian density (total Hamiltonian $\mathcal{H} = \int d^3\bm{x}~\mathscr{H}$)
\begin{equation}
    \mathscr{H} = \frac{\hbar^2}{2m} \nabla\Psi^\dagger(\bm x)\cdot \nabla\Psi(\bm x) + \frac{U}{2}\Psi^\dagger(\bm x)\Psi^\dagger(\bm x)\Psi(\bm x)\Psi(\bm x)\,,
\end{equation}
where $U>0$ is the 2-body contact repulsion.
Here $\Psi$ is the field operator, satisfying $[\Psi({\bm x}), \Psi^\dagger({\bm x'})] = \delta(\bm{x}- \bm{x}')$. 
The total momentum, in turn, reads $\bm{P} = - i \hbar \int d^3 {\bm x}\, \Psi^\dagger({\bm x}) \nabla \Psi(\bm{x})$.
From Eq.~\eqref{eq:dtAofx}, one finds the evolution of the expectation value of the field operator:
\begin{equation}\label{eq:grosspitaevskii}
    i\hbar\, \partial_t \langle\Psi\rangle = -\frac{\hbar^2}{2m}\,\nabla^2\langle\Psi\rangle + U\, \langle \Psi^\dagger \Psi \Psi\rangle + \frac{i\hbar^3\sigma^2}{2} \nabla^2\langle\Psi\rangle\,
    .
\end{equation}
We consider the mean-field approximation, where $\Psi$ is treated as a c-number function. 
This leads to a modified Gross-Pitaevskii equation~\cite{pethick_smith_2008}, which now contains an additional diffusion term.
We investigate the behavior of excitations on top of a homogeneous condensate, by defining $\Psi = \sqrt{n}\,e^{-i\mu t/\hbar}+\delta \Psi$, where $\mu = nU$ is the chemical potential, and $n$ is the background density (see Supplemental Material~\cite{SupMat}). 
To first order in $\delta \Psi$, and taking a Fourier 
transform $\delta \Psi \sim e^{-i (\mu/\hbar+\omega)t + i\bm k\cdot \bm x}$, we find the following dispersion relation for the excitations on top of the uniform background:
\begin{equation}
    \omega(k) = \pm k\,\sqrt{\frac{n\,U}{m} + \left(\frac{\hbar k}{2m}\right)^2 } -i\,\frac{\hbar^2 \sigma^2}{2}\,k^2\,.
\end{equation}
In the long wavelength limit, if $\sigma=0$, this reduces to the standard expression for a sound wave $\omega \sim c_s k$ in a superfluid, where $c_s = \sqrt{nU/m}$ \cite{pethick_smith_2008}. Conversely, when $\sigma \neq 0$, the long wavelength regime gives $\omega\thickapprox \pm c_s\,k-i \gamma_s k^2/2$. This expression has the same form as the Navier-Stokes dispersion relation for sound waves, where $\gamma_s=\hbar^2\sigma^2$ would then describe the attenuation of sound waves due to viscous effects \cite{landau2013fluid}.
Momentum dephasing therefore effectively adds viscosity to the superfluid, introducing  a dissipative Navier-Stokes-like contribution to the phonon dispersion relations.

\emph{\textbf{Additivity in linear response --}}
Arbitrarily far from equilibrium, Eq.~\eqref{eq:diffusion} generally predicts a complex interplay between unitary and diffusive contributions. Next, we show that within linear response, momentum dephasing leads to a universal positive contribution to the diffusion constant computed via a Kubo formula. 

Linear response theory (LRT) \cite{Nyquist1928,Callen1951,Onsager1953,Tisza1957,Kubo1966,Onsager1931,Onsager1931a} can be used to determine the diffusion constant including contributions from quantum noise (c.f. Ref.~\cite{Bertini2020b} for a recent review). Such LRT analyses for open quantum system have been carried out, e.g, in Refs.~\cite{Uchiyama2009,Uchiyama2010,Avron2011,Chetrite2012a,Ban2017,Mohammad2018,Konopik2018};  our approach differs from these, however, as we are specifically interested in the effects from momentum dephasing. The inclusion of momentum dephasing damps the expectation value of operators connecting different momentum eigenstates, producing entropy in the process. This is in contrast to the standard LRT treatment of closed quantum systems, where the von Neumann entropy is conserved.

Following \cite{forster1975hydrodynamic}, we assume that the system is initially in global equilibrium at $t = -\infty$, 
$\rho_\text{eq}=e^{-\beta \mathcal H+\beta \mu \mathcal{Q}}$, where $\beta=1/(k_B T)$, $\mu$ is the chemical potential, and $\mathcal{Q} = \int d^3 \bm{x}\, \mathcal{N}(\bm{x})$ is the total conserved charge. 
We then consider an adiabatic perturbation $\delta \mathcal{H}(t)$ of the system's Hamiltonian, of the form
\begin{equation}\label{pert}
\delta \mathcal{H}(t) =  -\int d^3 \bm{x}\,  \delta \mu(t,\bm{x}) \mathcal{N}(\bm{x}), 
\end{equation}
where $\delta \mu(t,\bm{x})$ represents inhomogeneities in the chemical potential. 
We assume that $\delta \mu(t,\bm{x}) = \theta(-t) e^{\epsilon t}  \delta \mu(\bm{x})$,
where $\epsilon >0$ is such that the perturbation slowly vanishes at $t\to -\infty$, and $\theta(-t)$ is the Heaviside step function which causes the perturbation to be abruptly turned off at $t=0$.

If the perturbations are small, we may expand $\rho(t) \simeq \rho_\text{eq} + \delta \rho$. 
Since $\mathcal{D} (\rho_\text{eq})=0$, inserting~(\ref{pert}) in~(\ref{eq:unperturbedLind}) we find that, to linear response in the perturbations, 
\begin{equation}\label{lrt_eq}
\frac{d \delta \rho}{d t} = \mathcal{L}(\delta \rho) - \frac{i}{\hbar} [\delta \mathcal{H}(t), \rho_\text{eq}], 
\end{equation}
where $\mathcal{L}$ is the Liouvillian~\eqref{eq:unperturbedLind}. 
The solution is 
\begin{equation}\label{lrt_sol}
\delta \rho(t) = -\frac{i}{\hbar} \int\limits_{-\infty}^t d t' \mathcal{E}_{t-t'} \bigg( [\delta \mathcal{H}(t'),\rho_\text{eq}] \bigg),
\end{equation}
where $\mathcal{E}_{t-t'} = e^{\mathcal{L}(t-t')}$ is the quantum channel associated with the unperturbed solution of Eq.~(\ref{eq:unperturbedLind}).

The deviation from equilibrium $\delta n(t,\bm{x}) := \tr\big\{ \mathcal{N}(\bm{x}) \delta \rho(t)\big\}$ can be written as 
\begin{equation}\label{lrt_green}
\delta n(t,\bm{x}) = \int\limits_{-\infty}^\infty d t' \int d^3\bm{x}'  \;G(t-t',\bm{x}-\bm{x}')\, \delta \mu(t',\bm{x}'),
\end{equation} 
where $G$ is the retarded Green's function,  given by
\begin{equation}\label{green}
G (t-t', \bm{x}- \bm{x}') = \frac{i}{\hbar} \theta(t-t')  \langle [\bar{\mathcal{E}}_{t-t'} (\mathcal{N}(\bm{x})), \mathcal{N} (\bm{x}') ] \rangle_\text{eq}.
\end{equation}
Here we also introduced the adjoint channel $\bar{\mathcal{E}}$, which is defined by the identity  $\tr\big\{\mathcal{O} \mathcal{E}(\rho)\big\} = \tr \big\{ \bar{\mathcal{E}} (\mathcal{O}) \rho \big\}$, for any operator $\mathcal{O}$. 
It can be shown that, since the master equation is of dephasing form,  $\bar{\mathcal{E}}$ can be obtained from the same evolution as~(\ref{eq:unperturbedLind}), but with $\mathcal{H} \to - \mathcal{H}$. 
As a consequence, from Eq.~\eqref{eq:unperturbedLind} it follows that
\begin{equation}\label{eq:NevoOp}
\partial_t \bar{\mathcal{E}}_t(\mathcal{N}(\bm{x})) = -\nabla \cdot  \bar{\mathcal{E}}_t (\bm{\mathcal{J}}(\bm x)) + \frac{\hbar^2\sigma^2}{2}\nabla^2\bar{\mathcal{E}}_t(\mathcal{N}(\bm{x})),
\end{equation}
where the momentum dephasing contributes both directly, in the second term, and indirectly, via the evolution of $\bm {\mathcal{J}}(\bm x)$. 

Differentiating~\eqref{lrt_green} with respect to time,  using~\eqref{eq:NevoOp}, and moving to Fourier space, one can show \cite{SupMat} that in the long wavelength limit the perturbation $\delta n(t, \bm{k})$ evolves according to 
\begin{IEEEeqnarray}{rCl}\label{eq:deltanavg}
    &&\partial_t\delta n(t,\bm k)= -D_{\textrm{deph}}\,k^2\delta n(t,\bm k) 
    \nonumber
    \\[0.2cm]
     && \quad
    + k^2\lim_{k\to 0}\frac{1}{k^2}\int\limits_{-\infty}^\infty dt' \partial_t G_{\textrm{uni}}(t-t',\bm{k}) \delta\mu(t',\bm{k}) + \mathcal{O}(k^4),
\end{IEEEeqnarray}
where $G_{\textrm{uni}}(t-t', \bm{k}) = G(t-t',\bm{k})\big|_{\sigma=0}$
is the  retarded Green's function associated with standard unitary linear response theory, under Hamiltonian $\mathcal{H}$~\cite{Bertini2020b,forster1975hydrodynamic}. 
This therefore shows that, as far as the transport coefficients are concerned,  the contribution from dephasing is additive. 
That is, the total diffusion constant can be written as 
\begin{equation}
 D = D_{\textrm{uni}} + D_{\textrm{deph}}\,,
 \label{totaldiffusion}
\end{equation}
where $D_{\textrm{uni}}$ is obtained from the standard  Kubo formula for $G_{\textrm{uni}}$ \cite{forster1975hydrodynamic}. In other words, in the long wavelength regime one finds 
\begin{equation}\label{eq:greenN}
 \lim\limits_{\omega\to 0}\lim\limits_{k \to 0} \frac{i\omega G(\omega,\bm k)}{k^2} =  \lim\limits_{\omega\to 0}\lim\limits_{k \to 0}  \frac{i\omega G_{\textrm{uni}}(\omega,\bm k)}{k^2}  +
\frac{\hbar^2\sigma^2}{2}\chi,
\end{equation}
where $\chi=(\partial n/\partial \mu)_T$ is the thermodynamic susceptibility. This shows that the total diffusion constant $D$ in \eqref{totaldiffusion}, which includes unitary and dephasing contributions, can be obtained by a Kubo formula.

A similar approach can also be used to calculate the  dephasing contribution to other transport coefficients, such as shear and bulk viscosities \cite{forster1975hydrodynamic}. In this case, we expect that momentum dephasing will also increase the values of those coefficients via the addition of a universal term $\sim \sigma^2$.

\emph{\textbf{Discussion --}} 
In this Letter we showed that the dephasing of the total momentum of a many-body system leads to diffusive behavior of macroscopic quantities, such as the system's particle density. 
We illustrated the emergence of diffusivity in three different ways. First, by showing that in the limit of strong dephasing one obtains a diffusion equation  even for states arbitrarily far from equilibrium. Second, by showing that dephasing damps sound waves in a superfluid, introducing a dissipative Navier-Stokes-like term to the dispersion relations of superfluid phonons. And third, in the context of linear response theory around equilibrium states, where we demonstrated that momentum dephasing enhances the diffusion constant of the system by a universal amount, associated solely to the strength of the coupling between the system and the  environment. A similar conclusion is expected to hold for shear and bulk viscosities. Another natural extension of our results can be done, for instance, by considering the dephasing of the total angular momentum.  

Dephasing has been shown to lead to diffusive behavior in the case of boundary-driven systems~\cite{Asadian2013,Znidaric2010a,landi2021nonequilibrium,Malouf2018,Znidaric2013a,Znidaric2017}. 
But this concerns dephasing of the local density operator, not momentum.
As we argued, due to charge conservation, dephasing of momentum is an expected feature of quantum many-body systems.
And while realistic system-bath interactions may generally lead to more complex momentum dephasing, the type studied here is shown to be particularly suited in capturing the long wavelength physics.

The emergence of a Navier-Stokes-like dispersion relation in a superfluid suggests that this effect may be directly tested experimentally using, e.g., ultracold atoms. 
Of course, the assumption of a noise that acts only in the total momentum is an idealization of the complex interactions between many-body systems and the environment.
However, our results  corroborate the general idea that decoherence should be deeply connected to the emergence of many macroscopic phenomena. 
Most notably, Eq.~\eqref{eq:diffusion} shows that diffusive behavior can emerge from decoherence even for systems arbitrarily far from thermodynamic equilibrium. 

Our results suggest that the range of applicability of hydrodynamics might be significantly broader than previously thought. This is in agreement with recent findings that hydrodynamic behavior may be applicable even far from equilibrium \cite{Florkowski:2017olj}, which is relevant for the quark-gluon plasma formed in ultrarelativistic heavy-ion collisions \cite{Heinz:2013th}. 
In fact, recently the framework of open quantum systems has been applied to a variety of phenomena in quantum chromodynamics involving heavy quarkonia, jets, and also the color glass condensate  
\cite{Akamatsu:2014qsa,DeBoni:2017ocl,Blaizot:2018oev,Brambilla:2020qwo,Yao:2021lus,Vaidya:2020lih,Vaidya:2020cyi,deJong:2020tvx,Armesto:2019mna}.
We hope our results may also contribute to such problems and shed some new light on the emergence of hydrodynamics in quantum many-body systems under extreme conditions.

\emph{\textbf{Acknowledgements --}} 
GTL acknowledges fruitful discussions with E.~Lutz. 
JN is partially supported by the U.S. Department of Energy, Office of Science, Office for Nuclear Physics under Award No. DE-SC0021301.
GTL acknowledges the financial support of the Gridanian Research Council (GRC), the  S\~ao Paulo Funding 
Agency FAPESP (Grants No. 2017/50304-7, 2017/07973-5
and 2018/12813-0) and the Brazilian funding agency CNPq
(Grant No. INCT-IQ 246569/2014-0).

\bibliography{library,other}

\onecolumngrid
\renewcommand{\theequation}{S.\arabic{equation}} 
\setcounter{equation}{0}
\pagebreak

\appendix
\begin{center}
    \huge Supplemental Material
\end{center}

Here we provide details on some of the main results of the paper. 
First, in section S.1. we describe the mean-field analysis of the superfluid problem and, in particular, the derivation of Eq.\ (10).
Then, in section S.2. we give additional details about the linear-response analysis.

\section{S.1. Dephasing viscosity}

We consider a superfluid with contact interactions described by the Hamiltonian density
\begin{equation}
    \mathscr{H}(\bm x) = \frac{\hbar^2}{2m} \nabla\Psi^\dagger(\bm x)\cdot \nabla\Psi(\bm x) + \frac{U}{2}\Psi^\dagger(\bm x)\Psi^\dagger(\bm x)\Psi(\bm x)\Psi(\bm x)\,,
\end{equation}
where $\Psi$ is the quantum wave-function operator. 
The evolution of the expectation value of  $\Psi(\bm x)$ can be obtained from Eq.\ (4) of the main text. 
From the commutation relation $[\Psi(\bm x),\Psi^\dagger(\bm x')]=\delta(\bm x-\bm x')$, one finds
\begin{equation}
    [\mathcal{H},\Psi(\bm x)] =\int d^3x' \,  [\mathscr{H}(\bm x'),\Psi(\bm x)] = \frac{\hbar^2}{2m} \nabla^2\Psi(\bm x) - U\,\Psi^\dagger(\bm x)\Psi^2(\bm x)\,.
\end{equation}
Combining this with the momentum dephasing $\mathcal{D}(\rho) = -(\sigma^2/2)\sum_k [P_k,[P_k,\rho]]$, yields
\begin{equation}\label{eq:grosspitaevskiisup}
    i\hbar\, \partial_t \langle\Psi\rangle = -\frac{\hbar^2}{2m}\,\nabla^2\langle\Psi\rangle + U\, \langle |\Psi|^2\Psi\rangle + \frac{i\hbar^3\sigma^2}{2} \nabla^2\langle\Psi\rangle\,.
\end{equation}

We investigate the behavior of excitations on top of a homogeneous condensate: $\Psi = \sqrt{n}\,e^{-i\mu t/\hbar}+\delta \Psi$. 
To order $\mathcal{O}(\delta\Psi^0)$, Eq.~\eqref{eq:grosspitaevskiisup} yields the constraint $\mu = n\,U$.  In the mean-field approximation and to first order in $\delta \Psi$, we obtain:
\begin{equation}\label{eq:evPsi}
    i\hbar \partial_t \delta \Psi = -\frac{\hbar^2}{2m}\,\nabla^2\delta\Psi+ \frac{i\hbar^3\sigma^2}{2} \nabla^2\delta\Psi+ 2n\,U\, \delta\Psi + n\,U\,e^{-2i\mu t/\hbar} \delta\Psi^\dagger \,.
\end{equation}
Taking the Fourier transform $ \delta \Psi(t,\bm x) = \int_{\omega}\int_{\bm k} \phi(\omega,\bm k)e^{-i (\mu+\omega)t + i\bm k\cdot \bm x}$, we find, from Eq.~\eqref{eq:evPsi} and its complex conjugate,  
\begin{equation}
 \left(\begin{array}{cc}
        \hbar \omega -\frac{\hbar^2\,k^2}{2m}(1-i\hbar\, m \,\sigma^2) - n\,U  &  n\,U\,e^{-2i\mu t/\hbar} \\
        n\,U\,e^{-2i\mu t/\hbar} & -\hbar \omega -\frac{\hbar^2\,k^2}{2m}(1+i\hbar\, m\, \sigma^2) - n\,U
    \end{array}\right) 
    \left(\begin{array}{c}
          \phi(\omega,\bm k)  \\
      \phi^*(\omega,\bm k)
    \end{array}
    \right) =
   \left(\begin{array}{c}
          0 \\
      0 
    \end{array}
    \right)\,.
\end{equation}
Finally, solving for $\omega$ yields the dispersion relation
\begin{equation}\label{eq:dispersion}
    \omega = \pm k\,\sqrt{\frac{n\,U}{m} + \left(\frac{\hbar k}{2m}\right)^2 } -i\,\frac{\hbar^2 \sigma^2}{2}\,k^2\,,
\end{equation}
which is Eq.\ (10) of the main text.

\section{S.2. Dephasing contribution to the diffusion constant}

Here, we present a detailed derivation of the dephasing contribution to the diffusion coefficient, obtained at long wavelength and low frequency (or long times). 
The  long-wavelength regime can be studied by moving 
to momentum space, by defining $\mathcal{A}(\bm{k}) = \int d^3 \bm{x} \; e^{i \bm{k} \cdot \bm{x}}\mathcal{A}(\bm{x})$, with $\mathcal{A}= \mathcal{N}, \bm{\mathcal{J}}$. 
From Eqs.~(13) and (16) in the main text, one finds
 \begin{equation}\label{eq:deltanavg}
    \partial_t\delta n(t,\bm k)= - \frac{\hbar^2\sigma^2}{2}k^2\delta n(t,\bm k) +\int_{-\infty}^t dt'\left\langle \left[ \bar{\mathcal{E}}_{t-t'}(\bm k\cdot\bm{ \mathcal{J}}(\bm k)),\mathcal{N}(\bm x=\bm 0) \right]\right\rangle_\textrm{eq}\,\delta\mu(t',\bm k),
\end{equation}
where $\delta n(t,\bm{x}) := \tr\big\{ \mathcal{N}(\bm{x}) \delta \rho(t)\big\}$.

The evolution of $\bm{\mathcal{J}}(\bm k)$ can be found from
\begin{equation}\label{eq:EbarEq}
\partial_t \bar{\mathcal{E}}_t(\bm{\mathcal{J}}(\bm{k})) = \frac{i}{\hbar}\bar{\mathcal{E}}_t (\left[\mathcal{H},\bm{\mathcal{J}}(\bm k)\right]) - \frac{\hbar^2\sigma^2}{2}k^2\,\bar{\mathcal{E}}_t(\bm{\mathcal{J}}(\bm{k})),
\end{equation}
which yields 
\begin{equation}
\partial_t \left(e^{\hbar^2\sigma^2k^2(t-t')/2}\bar{\mathcal{E}}_t(\bm{\mathcal{J}}(\bm{k})) \right)= \frac{i}{\hbar}\left[\mathcal{H},e^{\hbar^2\sigma^2k^2(t-t')/2}\bar{\mathcal{E}}_t (\bm{\mathcal{J}}(\bm k))\right] ,
\end{equation}
and the solution
\begin{equation}
    \bar{\mathcal{E}}_t(\bm{\mathcal{J}}(\bm{k})) = e^{-\hbar^2\sigma^2k^2(t-t')/2}\,U_{t-t'}^\dagger \bm{\mathcal{J}}(\bm{k}) U_{t-t'},
\end{equation}
where $U_{t-t'}$ is the unitary evolution operator.
Therefore,  Eq.~\eqref{eq:deltanavg} becomes
 \begin{eqnarray}\label{eq:unitarydecouples}
    \partial_t\delta n(t,\bm k)=&& - \frac{\hbar^2\sigma^2}{2}k^2\delta n(t,\bm k) +e^{-\hbar^2\sigma^2k^2(t-t')/2}\int_{-\infty}^t dt' \left\langle \left[ U_{t-t'}^\dagger\,\bm k\cdot\bm{ \mathcal{J}}(\bm k)\, U_{t-t'},\mathcal{N}(\bm x=\bm 0) \right]\right\rangle_\textrm{eq}\,\delta\mu(t',\bm x') \\
    =&& - \frac{\hbar^2\sigma^2}{2}k^2\delta n(t,\bm k) +e^{-\hbar^2\sigma^2k^2(t-t')/2}\int_{-\infty}^t dt'\, G_{\textrm{uni}}(t-t',\bm k)\,\delta\mu(t',\bm x') 
    \label{eq:unitaryGdecouples},
\end{eqnarray}
where we recognize the unitary (i.e., $\sigma=0$) Green's function $G_{\textrm{uni}}(t-t',\bm k):= G(t-t',\bm k)\big|_{\sigma=0}$ --- here, written in Fourier space, 
$G(t,\bm k)=\int d^3\bm x\, G(t,\bm x)\,e^{-i\bm k \cdot \bm x}$. 

In the long-wavelength limit, because the equilibrium state is isotropic, $\bm{\mathcal{J}}(\bm k)\sim \bm k$ in the average in Eq.~\eqref{eq:unitarydecouples}. 
Thus, in this limit, the $\sigma$ dependence of the second term in the rhs of Eq.~\eqref{eq:unitaryGdecouples} contributes only with subleading terms of $\mathcal{O}(\sigma^2 k^4)$. That is,
\begin{equation}\label{eq:deltanavgSup}
    \partial_t\delta n(t,\bm k)= -D_{\textrm{deph}}\,k^2\delta n(t,\bm k) 
    + k^2\lim_{k\to 0}\frac{1}{k^2}\int\limits_{-\infty}^\infty dt' \partial_t G_{\textrm{uni}}(t-t',\bm{k}) \,\delta\mu(t',\bm{k}) + \mathcal{O}(k^4,\sigma^2 k^4),
\end{equation}
where we already used that   $D_{\textrm{deph}}=\sigma^2\hbar^2/2$. 
Note that, for $\sigma=0$, the first term on the rhs vanishes so that,  in the appropriate limit, the second term must yield $-D_{\textrm{uni}}\,k^2\delta n(t,\bm k) $, where $D_{\textrm{uni}}$ is the unitary (i.e., $\sigma=0$) contribution to the diffusion coefficient.

Diffusive behavior sets in at long times. The long-time limit can be studied by taking the Fourier transform in time of the Green's function, $G(\omega,\bm k)=\int dt\, G(t,\bm k)\,e^{i\omega t}$. Also using that $\delta n(\omega,\bm k) = G(\omega,\bm k)\, \delta\mu(\omega,\bm k)$, Eq.~\eqref{eq:deltanavgSup} becomes
\begin{equation}
 \lim\limits_{k \to 0} \frac{i\omega G(\omega,\bm k)}{k^2} = 
  D_{\textrm{deph}}G(\omega,\bm k=\bm 0) 
 + \lim\limits_{k \to 0} \frac{i\omega G_{\textrm{uni}}(\omega,\bm k)}{k^2}.
\end{equation}
The long-time behavior corresponds to the low-frequency limit $\omega\to 0$:
\begin{equation}\label{eq:greenNsup}
 \lim\limits_{\omega\to 0}\lim\limits_{k \to 0} \frac{i\omega G(\omega,\bm k)}{k^2} = 
 \chi\,D_{\textrm{deph}}+
 \lim\limits_{\omega\to 0}\lim\limits_{k \to 0}  \frac{i\omega G_{\textrm{uni}}(\omega,\bm k)}{k^2} ,
\end{equation}
where $\chi=G(\omega=0,\bm k=\bm 0)=(\partial n/\partial \mu)_T$ is the thermodynamic susceptibility. 
Employing the Kubo formula for the diffusion constant \cite{forster1975hydrodynamic},
\begin{equation}
\chi D = - \lim\limits_{\omega \to 0} \lim\limits_{k \to 0}  \frac{\omega}{k^2} \text{Im}\,  G(\omega, \bm{k}),
\end{equation}
Eq.~\eqref{eq:greenNsup} yields 
\begin{equation}\label{eq:diffcoefsup}
     D = D_{\textrm{deph}} +  D_\textrm{uni},
\end{equation}
showing that momentum dephasing shifts the diffusion coefficient by a positive universal contribution. 

\end{document}